\documentclass{ws-ijmpd}
\usepackage{amsmath}
\usepackage{amssymb}
\usepackage{epsfig}
\usepackage{euscript}
\def\mathswitchr#1{\relax\ifmmode{\mathrm{#1}}\else$\mathrm{#1}$\fi}

\newcommand {\pslash}{\hbox{$\not\hbox{\kern-2.3pt $p$}$}}

\usepackage[compress]{cite}
\usepackage{epic}

\def\alf1{ {\alpha\over\pi} }

\begin{document}
\title{Prediction for the Cosmological Constant in Resummed Quantum Gravity and Constraints on SUSY GUT's}


\author{B.F.L. Ward}
\address{Physics Department, Baylor University, One Bear Place \# 97316\\
        Waco, TX 76798-7316, USA\\
        bfl\_ward@baylor.edu}

\maketitle
\begin{history}
\received{Day Month Year}
\revised{Day Month Year}
\end{history}

\begin{abstract}
We use our resummed quantum gravity approach to Einstein's general theory of relativity in the context of the Planck scale cosmology formulation of Bonanno and Reuter to estimate the value of the cosmological constant such that 
$\rho_\Lambda=(0.0024 eV)^4$. We argue that the closeness of this estimate to experiment constrains susy GUT models. We discuss in turn various theoretical issues that have been raised about the approach itself as well as about the application to estimate the cosmological constant. Given the closeness of the estimate to the currently observed value, we also discuss the theoretical uncertainty in the estimate -- at this time, we argue it is still large.
\end{abstract}

\ccode{04.60.Bc;04.62.+v;11.15.Tk}
\centerline{                BU-HEPP-14-09, ~Nov.,~ 2014}
\section{\bf Introduction}\label{intro}
\par
As Weinberg~\cite{wein1} has suggested,
the general theory of relativity may have a non-trivial UV fixed point, with a finite dimensional critical surface
in the UV limit. This would mean that it would be asymptotically safe~\cite{wein1} with an S-matrix
that depends on only a finite number of observable parameters. 
Strong evidence has been calculated in Refs.~\cite{reutera,laut,reuterb,reuter3,litim1,litim2,perc1,perc2,perc3,perc4,perc5},
using Wilsonian~\cite{kgw1,kgw2,kgw3,kgw4,kgw5,kgw6} field-space exact renormalization group methods, 
to support
Weinberg's asymptotic safety hypothesis for the Einstein-Hilbert theory.\par
In a parallel but independent development~\cite{bw1,bw2,bw2a,bw2b,bw2c,bw2d,bw2e,bw2f,bw2g,bw2h}, we have shown~\cite{bw2i} that the extension of the amplitude-based, exact resummation theory of Ref.~\cite{yfs11,yfs12, yfs-jw1,yfs-jw2,yfs-jw3,yfs-jw4,yfs-jw5,yfs-jw6,yfs-jw7,yfs-jw8,yfs-jw9,yfs-jw10,yfs-jw11,yfs-jw12,yfs-jw13,yfs-jw14} to the Einstein-Hilbert theory leads to UV-fixed-point behavior for the dimensionless
gravitational and cosmological constants. In our development,
we get the added bonus that the resummed theory is actually UV finite when expanded in the resummed propagators and vertices to any finite order in the respective improved loop expansion. We denote the attendant
resummed theory as {\em resummed quantum gravity}.\par 
In Ref.~\cite{ambj}\footnote{We also note that the model in Ref.~\cite{horva} realizes many aspects
of the effective field theory implied by the anomalous dimension of 2 at the
UV-fixed point but it does so at the expense of violating Lorentz invariance.}
more evidence for Weinberg's asymptotic safety behavior has been calculated using causal dynamical triangulated lattice methods.
At this writing, we can say that 
there is no known inconsistency between our analysis
and those of the Refs.~\cite{reutera,laut,reuterb,reuter3,litim1,litim2,perc1,perc2,perc3,perc4,perc5,ambj}.\par
We do note, however, that the results in Refs.~\cite{reutera,laut,reuterb,reuter3,litim1,litim2,perc1,perc2,perc3,perc4,perc5}, while impressive, involve cut-offs and  
some mild dependence on gauge parameters which remain in the results
to varying degrees even for products such as that for the UV limits of the 
dimensionless gravitational and cosmological constants. 
Accordingly, we refer to the approach in 
Refs.~\cite{reutera,laut,reuterb,reuter3,litim1,litim2,perc1,perc2,perc3,perc4,perc5} as the 
'phenomenological' asymptotic safety approach henceforward.
We would argue that the existence of the non-Gaussian UV 
fixed point found in these references is probably a physical result because
the respective 
dependencies are mild. Corroboration by a 
rigorously cut-off independent and gauge invariant calculation is required for 
the attendant results to be considered final. As our results are both gauge invariant and cut-off independent,
our approach offers such a calculation. Analogously, as 
the results from Refs.~\cite{ambj} involve 
lattice constant-type artifact 
issues, they are also only an indication of what the true continuum limit 
might realize. They need to be corroborated by a rigorous calculation 
without the issues of finite size and other possible lattice artifacts to 
be considered final. Our approach again offers an answer to these issues.
In view of the scenario just described, we can say that results of
Refs.~\cite{reutera,laut,reuterb,reuter3,litim1,litim2,perc1,perc2,perc3,perc4,perc5,ambj,horva} have prepared
the stage for us to try to fulfill the ultimate purpose of 
theoretical physics, that is to say, to make contact with experiment.\par
We proceed by observing that, in Refs.~\cite{reuter1,reuter2}, 
it has been argued that a realization\footnote{The attendant 
choice of the scale $k\sim 1/t$ used in Refs.~\cite{reuter1,reuter2} 
was also proposed in Ref.~\cite{sola1}.} of the successful
inflationary model~\cite{guth1,guth2,linde} of cosmology
may be provided by
the attendant phenomenological
asymptotic safety approach in 
Refs.~\cite{reutera,laut,reuterb,reuter3,litim1,litim2,perc1,perc2,perc3,perc4,perc5} 
to quantum gravity without the need of the as yet unseen inflaton scalar field: the attendant UV fixed point solution
allows one to develop Planck scale cosmology that joins smoothly onto
the standard Friedmann-Walker-Robertson classical descriptions. In this way
one arrives at a quantum mechanical 
solution to the horizon, flatness, entropy 
and scale free spectrum problems. In Ref.~\cite{bw2i}, we have shown
that, in the new
resummed theory~\cite{bw1,bw2,bw2a,bw2b,bw2c,bw2d,bw2e,bw2f,bw2g,bw2h} of quantum gravity, 
we not only recover the properties as used in Refs.~\cite{reuter1,reuter2} 
for the UV fixed point of quantum gravity but we also
get ``first principles''
predictions for the fixed point values of
the respective dimensionless gravitational and cosmological constants
in their analysis. 
In what follows here, we review how 
we carry~\cite{darkuni} the analysis one step further and arrive at an estimate
for the observed cosmological constant $\Lambda$ in the
context of the Planck scale cosmology of Refs.~\cite{reuter1,reuter2}.
We comment on the reliability of the result and its implications for susy GUTs
as well, as it will be seen
already to be close relatively to the observed value~\cite{cosm11,cosm12,pdg2008}.
Given the uncertainties that we will discuss, we will not
overdo the closeness to the experimental value. We will argue that this gives, at the least, some more credibility to the new resummed theory as well as to the methods in Refs.~\cite{reutera,laut,reuterb,reuter3,litim1,litim2,perc1,perc2,perc3,perc4,perc5,ambj}. We discuss more reflections on
the attendant implications of the latter credibility in the search for 
an experimentally testable union of the original ideas of Bohr and Einstein elsewhere~\cite{elswh}.\par
The discussion proceeds as follows. In the next section,
we recapitulate 
the Planck scale cosmology presented phenomenologically
in Refs.~\cite{reuter1,reuter2}. In Section 3,
we review our results in
Ref.~\cite{bw2i} for the dimensionless gravitational and cosmological constants
at the UV fixed point. In the course of this latter review, we give some of the elements  
the new proof in Ref.~\cite{darkuni} of the UV finiteness of the resummed quantum gravity
theory for the sake of completeness. In Section 4, we estimate 
the observed value of 
the cosmological constant $\Lambda$ by 
combining the Planck scale cosmology 
scenario in Refs.~\cite{reuter1,reuter2} with our results from the 
resummed quantum gravity theory. We discuss implications, limitations and criticisms of the estimate. Section 4 also contains our summary remarks.
\par
\section{\bf Recapitulation of Planck Scale Cosmology}

More specifically, let us recall the Einstein-Hilbert 
theory
\begin{equation}
{\cal L}(x) = \frac{1}{2\kappa^2}\sqrt{-g}\left( R -2\Lambda\right)
\label{lgwrld1a}
\end{equation} 
where $R$ is the curvature scalar, $g$ is the determinant of the metric
of space-time $g_{\mu\nu}$, $\Lambda$ is the cosmological
constant and $\kappa=\sqrt{8\pi G_N}$ for Newton's constant
$G_N$. The authors in Ref.~\cite{reuter1,reuter2} 
have used the phenomenological exact renormalization group
for the Wilsonian~\cite{kgw1,kgw2,kgw3,kgw4,kgw5,kgw6} coarse grained effective 
average action in field space to argue that
the attendant running Newton constant $G_N(k)$ and running 
cosmological constant
$\Lambda(k)$ approach UV fixed points as $k$ goes to infinity
in the deep Euclidean regime in the sense that 
$k^2G_N(k)\rightarrow g_*,\; \Lambda(k)\rightarrow \lambda_*k^2$
for $k\rightarrow \infty$ in the Euclidean regime.\par
To make the contact with cosmology the authors
in Refs.~\cite{reuter1,reuter2} use a phenomenological
connection between the momentum scale $k$ characterizing the coarseness
of the Wilsonian graininess of the average effective action and the
cosmological time $t$. In this way, the latter authors show~\cite{reuter1,reuter2} that the standard cosmological
equations admit of the following extension:
\begin{align}
(\frac{\dot{a}}{a})^2+\frac{K}{a^2}&=\frac{1}{3}\Lambda+\frac{8\pi}{3}G_N\rho\nonumber\\
\dot{\rho}+3(1+\omega)\frac{\dot{a}}{a}\rho&=0\nonumber\\
\dot{\Lambda}+8\pi\rho\dot{G_N}&=0\nonumber\\
G_N(t)&=G_N(k(t))\nonumber\\
\Lambda(t)&=\Lambda(k(t))
\label{coseqn1}
\end{align}
in a standard notation for the density $\rho$ and scale factor $a(t)$
with the Robertson-Walker metric representation as
\begin{equation}
ds^2=dt^2-a(t)^2\left(\frac{dr^2}{1-Kr^2}+r^2(d\theta^2+\sin^2\theta d\phi^2)\right)
\label{metric1}
\end{equation}
so that $K=0,1,-1$ correspond respectively to flat, spherical and
pseudo-spherical 3-spaces for constant time t.  When $p$ denotes the pressure,
the equation of state
is taken as 
\begin{equation} 
p(t)=\omega \rho(t).
\end{equation} 
The phenomenological functional relationship between the respective
momentum scale $k$ and the cosmological time $t$ is determined
in Refs.~\cite{reuter1,reuter2}
as
\begin{equation}
k(t)=\frac{\xi}{t}
\end{equation}
for some positive constant $\xi$ fixed
from requirements on
physically observable predictions.\par
The authors in Refs.~\cite{reuter1,reuter2}, using the UV fixed
points discussed above,
show that the system in (\ref{coseqn1}) admits, for $K=0$,
a solution in the Planck regime where $0\le t\le t_{\text{class}}$, with
$t_{\text{class}}$ a ``few'' times the Planck time $t_{Pl}$, which joins
smoothly onto a solution in the classical regime, $t>t_{\text{class}}$,
which coincides with standard Friedmann-Robertson-Walker phenomenology
but with the horizon, flatness, scale free Harrison-Zeldovich spectrum,
and entropy\footnote{Here, we should note that, to solve the entropy problem, the authors in Ref.~\cite{reuter2} retain the general form of the requirement from Bianchi's identity so that the second and third relations in (\ref{coseqn1}) are combined to
$\dot{\rho}+3(1+\omega)\frac{\dot{a}}{a}\rho = -
\frac{\dot{\Lambda}+8\pi\rho\dot{G_N}}{8\pi G_N}$; we discuss this in more detail in Sect. 4.} problems all solved purely by Planck scale quantum physics.\par
We note that 
the key properties of $g_*,\lambda_*$ used for the relevant analyses
in Refs.~\cite{reuter1,reuter2} 
are that the two UV limits are both positive and that the product 
$g_*\lambda_*$ is only mildly cut-off/threshold function dependent
 insofar as the details of the respective  Wilsonian coarse-graining procedure
are concerned. Accordingly,
here, we review the predictions in Refs.~\cite{bw2i} for these
UV limits derived from the resummed quantum gravity theory as presented in
~\cite{bw1,bw2,bw2a,bw2b,bw2c,bw2d,bw2e,bw2f,bw2g,bw2h} 
and we review how to use them to predict~\cite{darkuni} the current value of $\Lambda$.
In the interest of making the discussion self-contained and in view of the lack of familiarity of the resummed quantum gravity theory,
we start the next section
with a review of its basic principles. 
\par
\section{\bf $g_*$ and $\lambda_*$ in Resummed Quantum  Gravity -- A Review}
We start with the prediction for $g_*$, which we already presented in Refs.~\cite{bw1,bw2,bw2a,bw2b,bw2c,bw2d,bw2e,bw2f,bw2g,bw2h,bw2i}. Given that
the theory we use is still not very familiar, in the interest of completeness we recapitulate
the main steps in the calculation.
\par
Given that in the infrared regime which we shall
resum the graviton couples to a an elementary particle 
independently of the particle's spin, we may use a scalar
field to develop the required calculational framework.  
Accordingly, we start with the Lagrangian density for
the basic scalar-graviton system
which was considered by Feynman
in Refs.~\cite{rpf1,rpf2}:
\begin{equation}
\begin{split}
{\cal L}(x) &= -\frac{1}{2\kappa^2} R \sqrt{-g}
            + \frac{1}{2}\left(g^{\mu\nu}\partial_\mu\varphi\partial_\nu\varphi - m_o^2\varphi^2\right)\sqrt{-g}\\
            &= \quad \frac{1}{2}\left\{ h^{\mu\nu,\lambda}\bar h_{\mu\nu,\lambda} - 2\eta^{\mu\mu'}\eta^{\lambda\lambda'}\bar{h}_{\mu_\lambda,\lambda'}\eta^{\sigma\sigma'}\bar{h}_{\mu'\sigma,\sigma'} \right\}\\
            & \qquad + \frac{1}{2}\left\{\varphi_{,\mu}\varphi^{,\mu}-m_o^2\varphi^2 \right\} -\kappa {h}^{\mu\nu}\left[\overline{\varphi_{,\mu}\varphi_{,\nu}}+\frac{1}{2}m_o^2\varphi^2\eta_{\mu\nu}\right]\\
            & \quad - \kappa^2 \left[ \frac{1}{2}h_{\lambda\rho}\bar{h}^{\rho\lambda}\left( \varphi_{,\mu}\varphi^{,\mu} - m_o^2\varphi^2 \right) - 2\eta_{\rho\rho'}h^{\mu\rho}\bar{h}^{\rho'\nu}\varphi_{,\mu}\varphi_{,\nu}\right] + \cdots \\
\end{split}
\label{eq1-1}
\end{equation}
Here,
$\varphi(x)$ can be identified as the physical Higgs field as
our representative scalar field for matter,
$\varphi(x)_{,\mu}\equiv \partial_\mu\varphi(x)$,
and $g_{\mu\nu}(x)=\eta_{\mu\nu}+2\kappa h_{\mu\nu}(x)$
where we follow Feynman and expand about Minkowski space
so that $\eta_{\mu\nu}=diag\{1,-1,-1,-1\}$.
Continuing to follow Feynman, we have introduced the notation, 
$\bar y_{\mu\nu}\equiv \frac{1}{2}\left(y_{\mu\nu}+y_{\nu\mu}-\eta_{\mu\nu}{y_\rho}^\rho\right)$ for any tensor $y_{\mu\nu}$\footnote{Our conventions for raising and lowering indices in the 
second line of (\ref{eq1-1}) are the same as those
in Ref.~\cite{rpf2}.}.
The bare(renormalized) mass of our otherwise free BEH~\cite{beh1,beh2,beh3,beh4,atlas-cms-2012a,atlas-cms-2012b,atlas-cms-2012c,atlas-cms-2012d} field is $m_o$($m$) 
and we momentarily set the small
observed~\cite{cosm11,cosm12,pdg2008} value of the cosmological constant
to zero so that our quantum graviton, $h_{\mu\nu}$, has zero rest mass.
When we discuss phenomenology, we return to the latter point.
Feynman~\cite{rpf1,rpf2} has essentially worked out the Feynman rules for (\ref{eq1-1}), including the rule for the famous
Feynman-Faddeev-Popov~\cite{rpf1,ffp1a,ffp1b} ghost contribution needed 
for unitarity with the fixing of the gauge
(we use the gauge of Feynman in Ref.~\cite{rpf1},
$\partial^\mu \bar h_{\nu\mu}=0$). For this latter
material we refer the reader to Refs.~\cite{rpf1,rpf2}. 
We may thus address now directly the quantum loop corrections
in the theory in (\ref{eq1-1}).
\par
\begin{figure}
\begin{center}
\epsfig{file=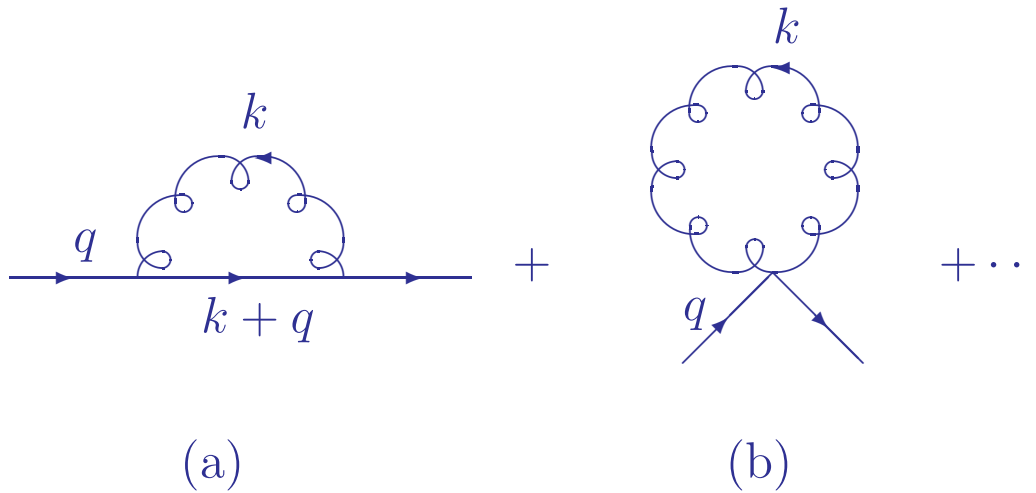,width=140mm}
\end{center}
\caption{\baselineskip=7mm     Graviton loop contributions to the
scalar propagator. $q$ is the 4-momentum of the scalar.}
\label{fig1}
\end{figure}

Referring to Fig.~\ref{fig1}, 
we have shown in Refs.~\cite{bw1,bw2,bw2a,bw2b,bw2c,bw2d,bw2e,bw2f,bw2g,bw2h} that the large virtual IR effects
in the respective loop integrals for 
the scalar propagator in quantum general relativity 
can be resummed to the {\em exact} result
\begin{equation}
\begin{split}
i\Delta'_F(k)&=\frac{i}{k^2-m^2-\Sigma_s(k)+i\epsilon}\cr
&=  \frac{ie^{B''_g(k)}}{k^2-m^2-\Sigma'_s+i\epsilon}\cr
&\equiv i\Delta'_F(k)|_{\text{resummed}}
\end{split}
\label{resum}
\end{equation}
for{\small ~~~($\Delta =k^2 - m^2$)
\begin{equation}
\begin{split} 
B''_g(k)&= -2i\kappa^2k^4\frac{\int d^4\ell}{16\pi^4}\frac{1}{\ell^2-\lambda^2+i\epsilon}\\
&\qquad\frac{1}{(\ell^2+2\ell k+\Delta +i\epsilon)^2}\\
&=\frac{\kappa^2|k^2|}{8\pi^2}\ln\left(\frac{m^2}{m^2+|k^2|}\right),       
\end{split}
\label{yfs1} 
\end{equation}}
where the latter form holds for the UV(deep Euclidean) regime, 
so that (\ref{resum}) 
falls faster than any power of $|k^2|$ -- by Wick rotation, the identification
$-|k^2|\equiv k^2$ in the deep Euclidean regime gives 
immediate analytic continuation to the result in the last line of (\ref{yfs1})
when the usual $-i\epsilon,\; \epsilon\downarrow 0,$ is appended to $m^2$. An analogous result~\cite{bw1} holds
for m=0; we refer the reader to Appendix 1 in Ref.~\cite{darkuni} 
where this is shown explicitly. Here, $i\Delta'_F(k)$ is the exact scalar propagator as
$-i\Sigma_s(k)$ is the 1PI scalar self-energy function.  
$\Sigma'_s$ starts in ${\cal O}(\kappa^2)$ so that
we may drop it in calculating one-loop effects. An important
consequence of the behavior of (\ref{resum}) is that,
when the respective analogs of (\ref{resum}) are used for the
elementary particles, one-loop 
corrections are finite. A stronger result actually holds~\cite{darkuni}: the use of
our resummed propagators renders all quantum 
gravity loops UV finite~\cite{bw1,bw2,bw2a,bw2b,bw2c,bw2d,bw2e,bw2f,bw2g,bw2h}. We have called this representation
of the quantum theory of general relativity resummed quantum gravity (RQG).
\par
It is important to understand that (\ref{resum}) is not limited to the regime where $k^2\cong m^2$
but is an identity that holds for all $k^2$. 
This can be demonstrated as follows.
If one inverts both sides of (\ref{resum}) 
one gets
\begin{equation}
\Delta_F^{-1}(k)-\Sigma_s(k)=(\Delta_F^{-1}(k)-\Sigma'_s(k))e^{-B''_g(k)}
\label{exact1}
\end{equation}
where the free inverse propagator is $\Delta_F^{-1}(k)=\Delta(k)+i\epsilon$.
We introduce here the loop expansions 
\begin{equation}
\Sigma_s(k)=\sum_{n=1}^{\infty}\Sigma_{s,n}(k),
\label{exact3a}
\end{equation}
\begin{equation}
\Sigma'_s(k)=\sum_{n=1}^{\infty}\Sigma'_{s,n}(k)
\label{exact3b}
\end{equation}
and, from elementary algebra, we get the exact relation
\begin{equation}
-\Sigma_{s,n}(k)=-\sum_{j=0}^{n}\Sigma'_{s,j}(k)\left(-B''_g(k)\right)^{n-j}/(n-j)!
\label{exact2}
\end{equation}
where we define for convenience $-\Sigma_{s,0}(k)=-\Sigma'_{s,0}(k)=\Delta_F^{-1}(k)$ and ${\cal A}_{s,n}$ is the n-loop contribution to ${\cal A}_s$. 
In this way we prove that every Feynman diagram contribution to $\Sigma_s(k)$ corresponds to a unique contribution to $\Sigma'_s(k)$ to all orders in $\kappa^2/(4\pi)$ for all values of $k^2$. QED.\par
An important issue to be resolved 
is whether the terms which we have extracted from 
the Feynman series in (\ref{exact2}) 
were actually in that series. In the limit
that $k^2\rightarrow m^2$, the result is known to be valid from
the arguments in Ref.~\cite{wein-qft} where the same result for the
respective exponentiating
virtual infrared divergence in (\ref{yfs1}) is obtained. 
More specifically, one generally 
introduces a regulator for the IR divergence and shows that
the terms which diverge as the regulator vanishes exponentiate
in the factor $B''_g(k)$. When $k^2\ne m^2$, the IR divergence is regulated by
$\Delta(k)$, so that we can use $\Delta(k)$ as our IR regulator. We may then
isolate that part of the amplitude which diverges when $\Delta(k) \rightarrow 0$ when the UV divergences are themselves regulated, by n-dimensional methods~\cite{thft-vlt} for example, so that they remain finite in this limit.
We stress the following: to any finite order in the loop expansion,
when we impose 
a gauge invariant regulator
for the UV regime, all UV divergences are regulated to finite results. 
If we resum the IR
dominant terms in the resultant UV-regulated theory, that resummation is valid independent of whether or not
the theory is UV renormalizable, as the theory is finite 
order by order in the loop expansion in the UV
when the UV regulator is imposed independent of whether or not it is renormalizable. The renormalizability issue arises only if we remove the UV regulator.
What we show Ref.~\cite{darkuni} in establishing (\ref{resum}) is that, after the IR resummation, the UV regulator can be removed and the UV regime remains finite order by order in the loop expansion
after the IR resummation.\par 
There is close analogy between our use of IR resummation in the presence of n-dimensional UV regularization to study the UV limit
of quantum gravity and the use of exact Wilsonian coarse graining
in Refs.~\cite{reutera,laut,reuterb,reuter3,litim1,litim2,perc1,perc2,perc3,perc4,perc5} to arrive at an effective average action 
for any given scale $k$ which has both an IR cut-off for momentum 
scales much smaller than $k$ and a UV cut-off for momentum scales much larger than $k$ so that the resulting field-space renormalization group
equation is well-defined even for a non-renormalizable theory like quantum gravity. In both cases the UV limit can be studied by taking the UV limit of the resulting non-perturbative solution and in both cases the same result obtains: a non-Gaussian UV fixed point is found, as we present below.\par
As we have discussed in Refs.~\cite{bw1,darkuni} 
and as Weinberg has shown in Ref.~\cite{wein-qft},
the IR limit of the coupling of the graviton to a particle is independent of its spin. It follows that we get the same exponential behavior 
as that shown in (\ref{resum}) in the resummed propagator for all particles
in the Standard Model. More precisely,
working now with the
complete theory
\begin{equation}
{\cal L}(x) = \frac{1}{2\kappa^2}\sqrt{-g} \left(R-2\Lambda\right)
            + \sqrt{-g} L^{\cal G}_{SM}(x)
\label{lgwrld1}
\end{equation}
where $L^{\cal G}_{SM}(x)$ is SM Lagrangian written in diffeomorphism
invariant form as explained in Refs.~\cite{bw1,bw2a}, 
when we use our resummed propagator results for all the particles
in the SM Lagrangian and for the graviton itself
we show~\cite{darkuni} in the Refs.~\cite{bw1,bw2,bw2a,bw2b,bw2c,bw2d,bw2e,bw2f,bw2g,bw2h} that the denominator for the propagation of transverse-traceless
modes of the graviton becomes ($M_{Pl}$ is the Planck mass)
\begin{equation}
q^2+\Sigma^T(q^2)+i\epsilon\cong q^2-q^4\frac{c_{2,eff}}{360\pi M_{Pl}^2},
\label{dengrvprp}
\end{equation}
where we have defined
\begin{equation}
\begin{split}
c_{2,eff}&=\sum_{\text{SM particles j}}n_jI_2(\lambda_c(j))\\
         &\cong 2.56\times 10^4
\end{split}
\label{c2eff}
\end{equation}
with $I_2$ defined~\cite{bw1,bw2,bw2a,bw2b,bw2c,bw2d,bw2e,bw2f,bw2g,bw2h}
by
\begin{equation}
I_2(\lambda_c) =\int^{\infty}_0dx x^3(1+x)^{-4-\lambda_c x}
\end{equation}
and with $\lambda_c(j)=\frac{2m_j^2}{\pi M_{Pl}^2}$ and~\cite{bw1,bw2,bw2a,bw2b,bw2c,bw2d,bw2e,bw2f,bw2g,bw2h}
$n_j$ equal to the number of effective degrees of particle $j$.
The numerical value in (\ref{c2eff}) corresponds to the following SM
masses: for the now presumed three massive neutrinos~\cite{neut,neuta,neutb},
we estimate a mass at $\sim 3$ eV; for
the remaining members
of the known three generations of Dirac fermions
$\{e,\mu,\tau,u,d,s,c,b,t\}$, we use~\cite{pdg2002,pdg2002a,pdg2004}
$m_e\cong 0.51$ MeV, $m_\mu \cong 0.106$ GeV, $m_\tau \cong 1.78$ GeV,
$m_u \cong 5.1$ MeV, $m_d \cong 8.9$ MeV, $m_s \cong 0.17$ GeV,
$m_c \cong 1.3$ GeV, $m_b \cong 4.5$ GeV and $m_t \cong 174$ GeV and for
the massive vector bosons $W^{\pm},~Z$ we use the masses
$M_W\cong 80.4$ GeV,~$M_Z\cong 91.19$ GeV, respectively.
We set the BEH mass at $m_{BEH}\equiv m_H\cong 126$GeV, in view of the
recent observations from ATLAS and CMS~\cite{atlas-cms-2012a,atlas-cms-2012b,atlas-cms-2012c,atlas-cms-2012d}.
We note that (see the Appendix 1 in Ref.~\cite{darkuni}) when the
rest mass of particle $j$ is zero, such as it is for the photon and the gluon,
the value of $m_j$ turns-out to be
$\sqrt{2}$ times the gravitational infrared cut-off
mass~\cite{cosm11,cosm12,pdg2008}, which is $m_g\cong 3.1\times 10^{-33}$eV.
We further note that, from the
exact one-loop analysis of Ref.\cite{tHvelt1}, it also follows (see Appendix 2
in Ref.~\cite{darkuni})
that the value of $n_j$ for the graviton and its attendant ghost is $42$.
For $\lambda_c\rightarrow 0$, we have found the approximate representation
(see Appendix 3 in Ref.~\cite{darkuni})
\begin{equation}
I_2(\lambda_c)\cong \ln\frac{1}{\lambda_c}-\ln\ln\frac{1}{\lambda_c}-\frac{\ln\ln\frac{1}{\lambda_c}}{\ln\frac{1}{\lambda_c}-\ln\ln\frac{1}{\lambda_c}}-\frac{11}{6}.
\end{equation} 
From these results we identify (we use $G_N$ for $G_N(0)$) 
\begin{equation}
G_N(k)=G_N/(1+\frac{c_{2,eff}k^2}{360\pi M_{Pl}^2})
\end{equation}
and compute the UV limit $g_*$ as
\begin{equation}
g_*=\lim_{k^2\rightarrow \infty}k^2G_N(k^2)=\frac{360\pi}{c_{2,eff}}\cong 0.0442.
\end{equation}
This result has no threshold/cut-off effects in it and
is a pure property of the known world. \par
To arrive at our prediction for $\lambda_*$, we use the Euler-Lagrange
equations to get Einstein's equation as 
\begin{equation}
G_{\mu\nu}+\Lambda g_{\mu\nu}=-\kappa^2 T_{\mu\nu}
\label{eineq1}
\end{equation}
in a standard notation where $G_{\mu\nu}=R_{\mu\nu}-\frac{1}{2}Rg_{\mu\nu}$,
$R_{\mu\nu}$ is the contracted Riemann tensor, and
$T_{\mu\nu}$ is the energy-momentum tensor. Using
the representation $g_{\mu\nu}=\eta_{\mu\nu}+2\kappa h_{\mu\nu}$
with the flat Minkowski metric $\eta_{\mu\nu}=\text{diag}(1,-1,-1,-1)$
we may isolate $\Lambda$ in Einstein's 
equation (\ref{eineq1}) by evaluating
its VEV(vacuum expectation value of both sides). 
We employ, for any bosonic quantum field $\varphi$, 
the point-splitting definition\footnote{We need to stress that this is a definition of convenience and is {\em not} a regularization because the integral which we calculate in (\ref{lambscalar}) below it is UV finite with exponential damping in the UV. The definition is robust, the direction of approach to the origin can be chosen arbitrarily, and when its vacuum expectation value is taken it may be replaced with the standard path integral Feynman rule for the tadpole loop that it most certainly is to give the same result.}  (here, :~~: denotes normal ordering as usual)
\begin{equation}
\begin{split}
\varphi(0)\varphi(0)&=\lim_{\epsilon\rightarrow 0}\varphi(\epsilon)\varphi(0)\cr
&=\lim_{\epsilon\rightarrow 0} T(\varphi(\epsilon)\varphi(0))\cr
&=\lim_{\epsilon\rightarrow 0}\{ :(\varphi(\epsilon)\varphi(0)): + <0|T(\varphi(\epsilon)\varphi(0))|0>\}\cr
\end{split}
\end{equation}
where the limit $\epsilon\equiv(\epsilon,\vec{0})\rightarrow (0,0,0,0)\equiv 0$
is taken from a time-like direction respectively. It follows that 
a scalar makes the contribution to $\Lambda$ given by\footnote{We note the
use here in the integrand of $2k_0^2$ rather than the $2(\vec{k}^2+m^2)$ in Ref.~\cite{bw2i}, to be
consistent with $\omega=-1$~\cite{zeld} for the vacuum stress-energy tensor.}
\begin{equation}
\begin{split}
\Lambda_s&=-8\pi G_N\frac{\int d^4k}{2(2\pi)^4}\frac{(2k_0^2)e^{-\lambda_c(k^2/(2m^2))\ln(k^2/m^2+1)}}{k^2+m^2}\cr
&\cong -8\pi G_N[\frac{1}{G_N^{2}64\rho^2}],\cr
\label{lambscalar}
\end{split}
\end{equation} 
where $\rho=\ln\frac{2}{\lambda_c}$ and we have used the calculus
of Refs.~\cite{bw1,bw2,bw2a,bw2b,bw2c,bw2d,bw2e,bw2f,bw2g,bw2h}
as recapitulated in Appendices 2,3 in Ref.~\cite{darkuni}. 
The standard methods relative to equal-time (anti-)commutation 
relations algebra realizations
then show that a Dirac fermion contributes $-4$ times $\Lambda_s$ to
$\Lambda$. We then see that the deep UV limit of $\Lambda$, allowing $G_N(k)$
to run as we calculated, is given by
\begin{equation}
\begin{split}
\Lambda(k) &\operatornamewithlimits{\longrightarrow}_{k^2\rightarrow \infty} k^2\lambda_*,\cr
\lambda_*&=-\frac{c_{2,eff}}{2880}\sum_{j}(-1)^{F_j}n_j/\rho_j^2\cr
&\cong 0.0817
\end{split}
\end{equation} 
where $F_j$ is the fermion number of $j$, $n_j$ is the effective
number of degrees of freedom of $j$ and $\rho_j=\rho(\lambda_c(m_j))$.
We note that $\lambda_*$ is free of threshold/cut-off effects and is
a pure prediction of our known world. Our result shows that 
$\lambda_*$ would vanish
in an exactly supersymmetric theory.\par
Our calculated UV fixed-point, 
$(g_*,\lambda_*)\cong (0.0442,0.0817)$, may be compared with the estimates
in Refs.~\cite{reuter1,reuter2}, 
which give $(g_*,\lambda_*)\approx (0.27,0.36)$. 
Here, one must keep in mind that 
the analysis in Refs.~\cite{reuter1,reuter2} did not include
the specific SM matter action and that there is definitely cut-off function
sensitivity to the results in the latter analyses. What we stress
is that the qualitative results that $g_*$ and $\lambda_*$ are 
both positive and are less than 1 in size 
are true of our results as well.\par
If we restrict our resummed quantum gravity
calculations above for $g_*,\lambda_*$ to 
the pure gravity theory, we get the results
$$g_*=.0533,\;\lambda_*=-.000189.$$ These results suggest
that there are still significant cut-off effects in the results 
used for $g_*,\;\lambda_*$ in Refs.~\cite{reuter1,reuter2}. The latter
results already
seem to include an effective matter contribution when viewed from
our resummed quantum gravity perspective, as an artifact of the gauge and cut-off dependencies of the results. More sepcifically, 
from a purely quantum field theoretic point of view,
the cut-off action is 
\begin{equation}
\Delta_kS(h,C,\bar{C};\bar{g})=\frac{1}{2}<h,{\cal R}^{\text{grav}}_kh>+<\bar{C},{\cal R}^{\text{gh}}_kC>
\end{equation}
where $\bar{g}$ is the general background metric, which is the Minkowski space metric $\eta$ here, 
and $C,\bar{C}$ are the
ghost fields and the operators ${\cal R}^{\text{grav}}_k,\; {\cal R}^{\text{gh}}_k$ implement the course graining as they satisfy the limits 
\begin{equation}
\begin{split}
{\underset{p^2/k^2\rightarrow \infty}{\text{lim}}} {\cal R}_k &=0,\nonumber\\
{\underset{p^2/k^2\rightarrow 0}{\text{lim}}}{\cal R}_k&\rightarrow \mathfrak{Z}_k k^2,
\end{split}
\end{equation}
for some $\mathfrak{Z}_k$~\cite{laut}. Here, the inner product is that defined
in Ref.~\cite{laut} in its Eqs.(2.14,2.15,2.19).
The result is that the modes with $p\lesssim k$ have a shift of their vacuum energy
by the cut-off operator. It follows that there is 
no disagreement in principle between
our gauge invariant results and the gauge dependent and cut-off dependent results in Refs.~\cite{laut}.
In other words, the graviton and ghost fields
at low scales compared to k have a mass added to them, so that their vacuum energies are shifted by a mass of order k. This shows up as a positive
contribution to the cosmological constant and explains why the EFRG result
for $\lambda_*$ has a positive value in the regime of the gauge parameter
in Ref.~\cite{laut} where the UV fixed point is attractive.\par

\section{\bf Review of An Estimate of $\Lambda$~\cite{darkuni}}

To see that the results in the previous Section, taken together with those in Refs.~\cite{reuter1,reuter2}, allow us to estimate the value of $\Lambda$ today, we take the normal-ordered form of Einstein's equation 
\begin{equation}
:G_{\mu\nu}:+\Lambda :g_{\mu\nu}:=-\kappa^2 :T_{\mu\nu}: .
\label{eineq2}
\end{equation}
The coherent state representation of the thermal density matrix then gives
the Einstein equation in the form of thermally averaged quantities with
$\Lambda$ given by our result in (\ref{lambscalar}) summed over 
the degrees of freedom as specified above in lowest order. In Ref.~\cite{reuter2}, arguments are presented that the Planck scale cosmology description of inflation needs the transition time between the Planck regime and the classical Friedmann-Robertson-Walker(FRW) regime at $t_{tr}\sim 25 t_{Pl}$. (We comment below on the uncertainty of this choice of $t_{tr}$.)\footnote{The analysis in Ref.~\cite{reuter2} of their renormalization group
improved Einstein equations finds a set of solutions in which one has power law inflation in the UV regime and one switches
abruptly to the classical FRW solution with essentially zero cosmological constant at the transition time $t_{tr}$. In other words, 
the solution to the renormalization group improved Einstein equations at the transition time and later is very well approximated by non-running values of the gravitational and cosmological constant when one uses the FRW approximation. This also avoids issues of double counting of effects, for example. From our (\ref{eq-rho-expt}) one sees that allowing the running to continue past $t_{tr}$ would not change our result for $\rho_\Lambda$ by very much at all, less than 8\%. We ignore effects of such size here.}
Hence, we write
\begin{equation}
\begin{split}
\rho_\Lambda(t_{tr})&\equiv\frac{\Lambda(t_{tr})}{8\pi G_N(t_{tr})}\cr
         &=\frac{-M_{Pl}^4(k_{tr})}{64}\sum_j\frac{(-1)^Fn_j}{\rho_j^2}.
\end{split}
\label{eq-rho-lambda}
\end{equation}
We further use the arguments in Refs.~\cite{branch-zap1,branch-zap2} ($t_{eq}$ is the time of matter-radiation equality) to get, from the method of the operator field, the 
first principles estimate
\begin{equation}
\begin{split}
\rho_\Lambda(t_0)&\cong \frac{-M_{Pl}^4(1+c_{2,eff}k_{tr}^2/(360\pi M_{Pl}^2))^2}{64}\sum_j\frac{(-1)^Fn_j}{\rho_j^2}\cr
          &\qquad\quad \times \frac{t_{tr}^2}{t_{eq}^2} \times (\frac{t_{eq}^{2/3}}{t_0^{2/3}})^3\cr
    &\cong \frac{-M_{Pl}^2(1.0362)^2(-9.194\times 10^{-3})}{64}\frac{(25)^2}{t_0^2}\cr
   &\cong (2.4\times 10^{-3}eV)^4.
\end{split}
\label{eq-rho-expt}
\end{equation}
where we take the age of the universe to be $t_0\cong 13.7\times 10^9$ yrs. 
In the estimate in (\ref{eq-rho-expt}), the first factor in the second line comes from the period from
$t_{tr}$ to $t_{eq}$ which is radiation dominated and the second factor
comes from the period from $t_{eq}$ to $t_0$ which is matter dominated
\footnote{The method of the operator field forces the vacuum energies to follow the same scaling as the non-vacuum excitations.}.
The result (\ref{eq-rho-expt}) should be compared with the experimental result~\cite{pdg2008}\footnote{See also Ref.~\cite{sola2} for an analysis that suggests 
a value for $\rho_\Lambda(t_0)$ that is qualitatively similar to this experimental result.} 
$\rho_\Lambda(t_0)|_{\text{expt}}\cong ((2.37\pm 0.05)\times 10^{-3}eV)^4$. 
\par
To sum up, in addition to our having put the  
Planck scale cosmology~\cite{reuter1,reuter2} on a
more rigorous basis, we believe our result for 
$\rho_\Lambda(t_0)$ is an estimate that represents some amount of progress in
the long effort to understand its observed value  
in quantum field theory. Evidently, as hitherto unseen degrees of freedom 
may exist and they have not been included, for example, our estimate is not a precision prediction.\par
It is interesting, in view of the Appelquist-Carazzone decoupling theorem~\cite{ta-jc}, that our result for the contribution to $\Lambda$ from a particle of rest mass $m$ scales as $1/\ln^2(2/\lambda_c(m))$ 
so that for masses $m<<M_{Pl}$
the larger the mass, the larger the contribution in magnitude. Specifically, the t, b, c, s, d, u,  $\tau$, $\mu$, e and the three neutrinos (together) contribute respectively 21.1\%, 17.6\%, 16.7\%, 
15.2\% , 13.5\%, 13.2\%, 5.63\%, 4.97\%, 4.01\% and 7.93\% of $\Lambda$ whereas the Higgs, W and Z 
bosons contribute -1.73\%, -5.10\% and -10.1\% of $\Lambda$ respectively.
The photon and the gluon, taken together, contribute -2.51\% of $\Lambda$, while the graviton contributes -0.277\%
thereof. Naively, our results are unexpected since by the Appelquist-Carazzone decoupling theorem larger values of $m$ might be expected to be more suppressed. We comment as follows. First, recall that the decoupling theorem in Ref.~\cite{ta-jc} was not proved for (power-countingly) nonrenormalizable theories such as the Einstein-Hilbert theory we 
deal with here. Our resummation renders the theory UV finite
with a characteristic scale of $\sim M_{Pl}$ for the scale beyond 
which the UV modes are suppressed and this is again in contradiction with the hypothesis of the Appelquist-Carazzone theorem. Note that in
the analyses presented above, we assume that $m/M_{Pl}<<1$ in deriving our results. For the 
integral on the RHS of the (\ref{lambscalar}) for $\Lambda_s$, 
which diverges like 4-powers of the cut-off without resummation and which
has a dependence on $M_{Pl}^4$ when we resum the theory, 
the remaining dependence on the
particle mass $m$ arises from the strength of the suppression of the modes beyond the characteristic scale $M_{Pl}$. The suppression is stronger for the smaller values of $m$ as they are farther away from the dominant scale $M_{Pl}$,
in accordance with what we expect from the uncertainty principle. 
The situation becomes even more
transparent if we consider masses $m>>M_{Pl}$, so that we are not subject
to effects of finite physical intrinsic scales. For two masses $m_1,\; m_2$
with $m_i>> M_{Pl}$, the contribution to $\Lambda_s$
scales as $m_iM_{Pl} $. This is the behavior one would expect from
summing the zero modes of a field of rest mass $m_i$ when the resummation 
causes the phase space integral to cut-off at a scale $\sim M_{Pl}$ and thereby yields
the factor $-8\pi G_N(M_{Pl}^3m_i)$. This factor follows from the 
fact that the vacuum energy density of the field is
given by (Here ${\cal H}$ is the usual free field Hamiltonian density.)
$$<0|{\cal H}|0>\sim \int^{M_{Pl}}\frac{ d^3 k}{(2\pi)^3}\frac{1}{2} \omega(k)= \int^{M_{Pl}}\frac{ d^3 k}{(2\pi)^3} \frac{1}{2}\sqrt{k^2+m_i^2}$$
where $\omega(k)$ is the usual frequency for mode $\vec{k}$ of the field -- 
$\omega(k)$ reduces
to $m_i$ when $ k^2<<m_i^2$. Since its zero modes are larger, the larger mass makes a larger contribution. Accordingly, it seems prudent to consider
what would happen to our estimate in the presence of a GUT theory at high scale?We turn next to this.\par
In Ref.~\cite{darkuni} we considered the susy SO(10) GUT scenario in Ref.~\cite{ravi-1}. In this scenario, the break-down of the GUT gauge symmetry to the 
low energy gauge symmetry occurs with an intermediate stage with gauge group
$SU_{2L}\times SU_{2R}\times U_1\times SU(3)^c$ where the final break-down to the Standard Model~\cite{gsw1,gsw2,gsw3,gsw4,gsw5,gsw6,gsw7,qcd1,qcd2,qcd3} gauge group, $SU_{2L}\times U_1\times SU(3)^c$, occurs at a scale $M_R\gtrsim 2TeV$ while the breakdown of global susy occurs at the (EW) scale $M_S$ which satisfies $M_R > M_S$.
Note that only the broken susy multiplets can contribute to the RHS of
(\ref{eq-rho-lambda}). In Ref.~\cite{darkuni}, we take, in view of the recent LHC results~\cite{lhc-susy},  
for illustration the values $M_R\cong 4 M_S\sim 2.0{\text{TeV}}$ and set the following susy partner values:
\begin{equation}
\begin{split}
m_{\tilde{g}}&\cong 1.5(10){\text{TeV}}\\
m_{\tilde{G}}&\cong 1.5{\text{TeV}}\\
m_{\tilde{q}}&\cong 1.0{\text{TeV}}\\
m_{\tilde{\ell}}&\cong 0.5{\text{TeV}}\\
m_{\tilde{\chi}^0_i}&\cong\begin{cases} &0.4{\text{TeV}},\;i=1\\
                                        & 0.5{\text{TeV}},\; i=2,3,4
                    \end{cases}\\
m_{\tilde{\chi}^{\pm}_i}&\cong  0.5{\text{TeV}},\; i=1,2\\
m_{S}&= .5{\text{TeV}},\; S=A^0,\; H^{\pm},\; H_2,
\end{split}
\end{equation}  
where we use a standard notation for the susy partners of the known quarks($q\leftrightarrow \tilde{q}$), leptons($\ell\leftrightarrow \tilde{\ell}$) and gluons($G\leftrightarrow \tilde{G}$), and the EW gauge and Higgs bosons($\gamma,\; Z^0,\; W^{\pm},\;H,$
$A^0,\;H^{\pm},\;H_2  \leftrightarrow \tilde{\chi}$)  with the extra Higgs particles denoted as usual~\cite{haber} by $A^0$(pseudo-scalar), $H^{\pm}$(charged) and $H_2$(heavy scalar). $\tilde{g}$ is the gravitino, for which we show two examples of its mass for illustration. As we discuss in Ref.~\cite{darkuni},
these particles generate extra contributions to the sum in the first line
on the RHS of 
(\ref{eq-rho-expt}) which lead to the values
$\rho_\Lambda = -(1.67\times 10^{-3}\text{eV})^4(-(1.65\times 10^{-3}\text{eV})^4)$, respectively, for the two values of $m_{\tilde g}$.
The positive observed value quoted above would appear to be in conflict with the sign of these results by many standard deviations, even if one allows for the considerable uncertainty in the various other factors in (\ref{eq-rho-expt}), which are all positive in our framework. As we note in Ref.~\cite{darkuni}, 
this may be alleviated 
in two ways. In approach (A), we may add new particles to the model. In approach (B), we allow a soft susy breaking mass term for the gravitino that resides near the GUT scale~\cite{ravi-1}
$M_{GUT}\sim 4\times 10^{16} GeV$. In approach (A), while doubling the number of quarks and leptons, we invert
the mass hierarchy between susy partners, so that the new squarks and sleptons are lighter than the new quarks and leptons. This works as long as
as we increase $M_R,\; M_S$ and have the new quarks and leptons
at $M_{\text{High}}\sim 3.4(3.3)\times 10^3\text{TeV}$ while leaving their partners at $M_{\text{Low}}\sim .5{\text{TeV}}$. For approach (B), the mass of the gravitino soft breaking term is set to
$m_{\tilde{g}}\sim 2.3\times 10^{15}{\text{GeV}}$. To summarize, we see that our
estimate in (\ref{eq-rho-expt}) can be used
as a constraint of general susy GUT models which we hope 
to explore in more detail elsewhere.\par   
As we have explained in Ref.~\cite{darkuni}
the value of $t_{tr}$ cannot be taken as precise. Specifically,
we argue~\cite{darkuni} that theory 
we are using for it from Ref.~\cite{reuter2} is uncertain by
a couple of orders of magnitude, which translates into an uncertainty of 
$\sim 10^4$ on
our estimate of $\rho_\Lambda$.\par
Moreover, we have addressed in Ref.~\cite{darkuni}
three other important matters that we have not mentioned:(1), the effect of the various spontaneous symmetry vacuum energies on our 
$\rho_{\Lambda}$ estimate methodology as exhibited here; (2), the issue of the impact of our approach on big bang nucleosynthesis(BBN)~\cite{bbn}; and, (3), the covariance of theory in the presence of time dependent values of $\Lambda$ and of $G_N$. What we show in Ref.~\cite{darkuni} is that 
the various spontaneous symmetry vacuum energies have little effect on our 
$\rho_{\Lambda}$ estimate methodology because of their small relative size to the Planck mass $M_{\text{Pl}}$ and that our approach has negligible effect on the standard BBN. In addition, as explained in Ref.~\cite{darkuni}, while 
only when $\dot{\Lambda}+8\pi\rho \dot{G}_N=0$ holds is covariant conservation of matter in the current universe guaranteed and while either the case with or the case without such guaranteed conservation is possible provided the attendant deviation is small(Detailed studies
of such deviation, including its maximum possible size, can be found in Refs.~\cite{bianref1,bianref2,bianref3}.), the results in this Section, unlike
(\ref{coseqn1}), use 
the more general realization of the attendant 
Bianchi identity requirement, from which
we have \begin{equation}\dot{\rho}+3\frac{\dot{a}}{a}(1+\omega)\rho= -\frac{\dot{\Lambda}+8\pi\rho \dot{G}_N}{8\pi G_N}\end{equation}
to be compared with (\ref{coseqn1}).\par
Finally, let us comment on an apparent 
ongoing confusion about the UV limit of the
Einstein-Hilbert theory we tame here
and the infrared limits of the same theory 
discussed in Refs.~\cite{dono1,dono2,bjrrm}. 
We stress that we have exponentiated the 
terms $\frac{\kappa^2|k^2|}{8\pi^2}
\ln\left(\frac{\mu^2}{|k^2|}\right)$ in $B''_g$ for a massless graviton 
in the deep UV for $|k^2|\rightarrow \infty$
which Donoghue
{\it et al.} and Bjerrum-Bohr treat to leading order for $k^2\rightarrow 0$ for the large distance limit. Accordingly, we are completely consistent with what Donoghue {\it et al.} and Bjerrum-Bohr have found in their respective analyses.
\par
As we noted, the gauge invariance of Feynman's formulation of Einstein's theory has been emphasized by Feynman himself in Ref.~\cite{rpf1,rpf2}. Here, we observe the following for the sake of completeness. The infrared exponent that we resum is gauge invariant as it is the most singular contribution
to one-loop 1PI 2-point proper vertex function at the infrared point and any infinitesimal diffeomorphism transformation adds gradient tensors to the graviton field $h_{\mu\nu}$ in Feynman's notation all of which vanish at the infrared point. Since the S-matrix is gauge invariant and we make exact rearrangement with a gauge invariant exponent, our resummed theory is also gauge invariant.\par
We sum up as follows. One must note that 
the model Planck scale cosmology of Bonanno and Reuter
which we use is just that, a model. More work needs
to be done to remove 
from it the type of uncertainties which we just elaborated in our estimate of $\Lambda$. With this latter goal in mind,
we do look forward to additional possible checks from observations. \par
\section*{Acknowledgments}
We thank Profs. L. Alvarez-Gaume and W. Hollik for the support and kind
hospitality of the CERN TH Division and the Werner-Heisenberg-Institut, MPI, Munich, respectively, where a part of this work was done.
\par
\section*{Note Added:}
Here, we point out for clarity that in computing $\Lambda$
in the Planck regime the assumption of $K=0$ is presumed 
as that is the only case for which the Bonanno-Reuter Planck scale
cosmology has been shown to allow a smooth connection from
the Planck regime for times near or earlier than the Planck time
to the semi-classical FRW regime for times after $t_{tr}$.
For $K=0$, by definition, equal time slices are flat 3-spaces, exactly
as we have employed in the vacuum states used to compute 
the zero-point energies that comprise $\Lambda$. Thus the results
in Sections 3 and 4 are fully self-consistent.

\end{document}